\documentclass{PoS}
\usepackage{amsmath,amssymb}
\usepackage{slashed}
\usepackage{cite}
\newcommand{\lb}{\left(}
\newcommand{\rb}{\right)}
\newcommand{\lam}{\lambda}
\newcommand{\GeV}{{\ensuremath\rm GeV}}
\newcommand{\TeV}{{\ensuremath\rm TeV}}
\newcommand{\fb}{{\ensuremath\rm fb}}
\newcommand{\ab}{{\ensuremath\rm ab}}

\newcommand{\be}{\beta}
\newcommand{\al}{\alpha}
\newcommand{\eqn}{equation}

\title{The IDM and THDMa - current constraints and future prospects}
 \ShortTitle{IDM and THDMa}
\author{\speaker{Tania Robens}\\
        Division of Theoretical Physics, Ruder Boskovic Institute, Zagreb, Croatia\\
        E-mail: \email{trobens@irb.hr}}

\abstract{We discuss two models that extend the Standard Model by additional particles, leading to additional scalar states, and also providing dark matter candidates. We briefly review the current constraints and comment on production cross sections at present and future collider facilities.\\ RBI-ThPhys-2021-36}

\FullConference{%
  *** The European Physical Society Conference on High Energy Physics (EPS-HEP2021), ***\\
  *** 26-30 July 2021 ***\\
  *** Online conference, jointly organized by Universit\"at Hamburg and the research center DESY ***
}

\begin{document}
\maketitle

\section{Introduction}
We discuss two new physics models that extend the Standard Model (SM) particle sector by additional scalars and also both provide a dark matter candidate. The models are confronted with current theoretical and experimental constraints. From the theoretical side, these include the minimization of the vacuum as well as the requirement of vacuum stability and positivity. We also require perturbative unitarity to hold, and perturbativity of the couplings at the electroweak scale.

Experimental bounds include the agreement with current measurements of the properties of the 125 \GeV~ resonance discovered by the LHC experiments, as well as agreement with the null-results from searches for additional particles at current or past colliders. We also confront the models with bounds from electroweak precision observables via $S,\,T,\,U$ \cite{Altarelli:1990zd,Peskin:1990zt,Peskin:1991sw}, B-physics observables $\lb B\,\rightarrow\,X_s\,\gamma,\,B_s\,\rightarrow\,\mu^+\,\mu^-,\,\Delta M_s\rb$, as well as agreement with astrophysical observables (relic density and direct detection bounds). We use a combination of private and public tools in these analyses, where the latter include HiggsBounds \cite{Bechtle:2020pkv},  HiggsSignals \cite{Bechtle:2020uwn}, 2HDMC \cite{Eriksson:2009ws}, SPheno \cite{Porod:2011nf}, Sarah \cite{Staub:2013tta}, micrOMEGAs \cite{Belanger:2018ccd,Belanger:2020gnr}, and MadDM \cite{Ambrogi:2018jqj}. Experimental numbers are taken from \cite{Baak:2014ora,Haller:2018nnx} for electroweak precision observables, \cite{combi} for $B_s\,\rightarrow\,\mu^+\,\mu^-$, \cite{Amhis:2019ckw} for $\Delta M_s$ and \cite{Planck:2018vyg} and  \cite{Aprile:2018dbl} for relic density and direct detection, respectively.
Bounds from $B\,\rightarrow\,X_s\gamma$ are implemented using a fit function from \cite{Misiak:2020vlo,mm}. Predictions for production cross sections shown here have been obtained using  Madgraph5 \cite{Alwall:2011uj}.
\section{Inert Doublet Model}

The Inert Doublet Model (IDM) \cite{Deshpande:1977rw,Cao:2007rm,Barbieri:2006dq} is a two-Higgs-doublet Model (THDM) that obeys an exact $Z_2$ symmetry, under which the SM-like particles are even, while the novel scalars from the second (dark) doublet are odd. The exactness of the symmetry induces directly stability of the lightest new scalar state, providing a dark matter candidate. The model has been widely discussed in the literature, see e.g. \cite{Ilnicka:2015jba,Kalinowski:2018ylg,Dercks:2018wch,Kalinowski:2020rmb}. 
The dark sector  {contains} four new particles: $H$, $A$ and $H^{\pm}$. After electroweak symmetry breaking, the model contains 5 free parameters which we chose to be $M_H, M_A, M_{H^{\pm}}, \lam_2, \lam_{345}\,\equiv\,\lam_3+\lam_4+\lam_5$.

\begin{center}
\begin{figure}
\begin{center}
\begin{minipage}{0.42\textwidth}
\begin{center}
\includegraphics[width=\textwidth]{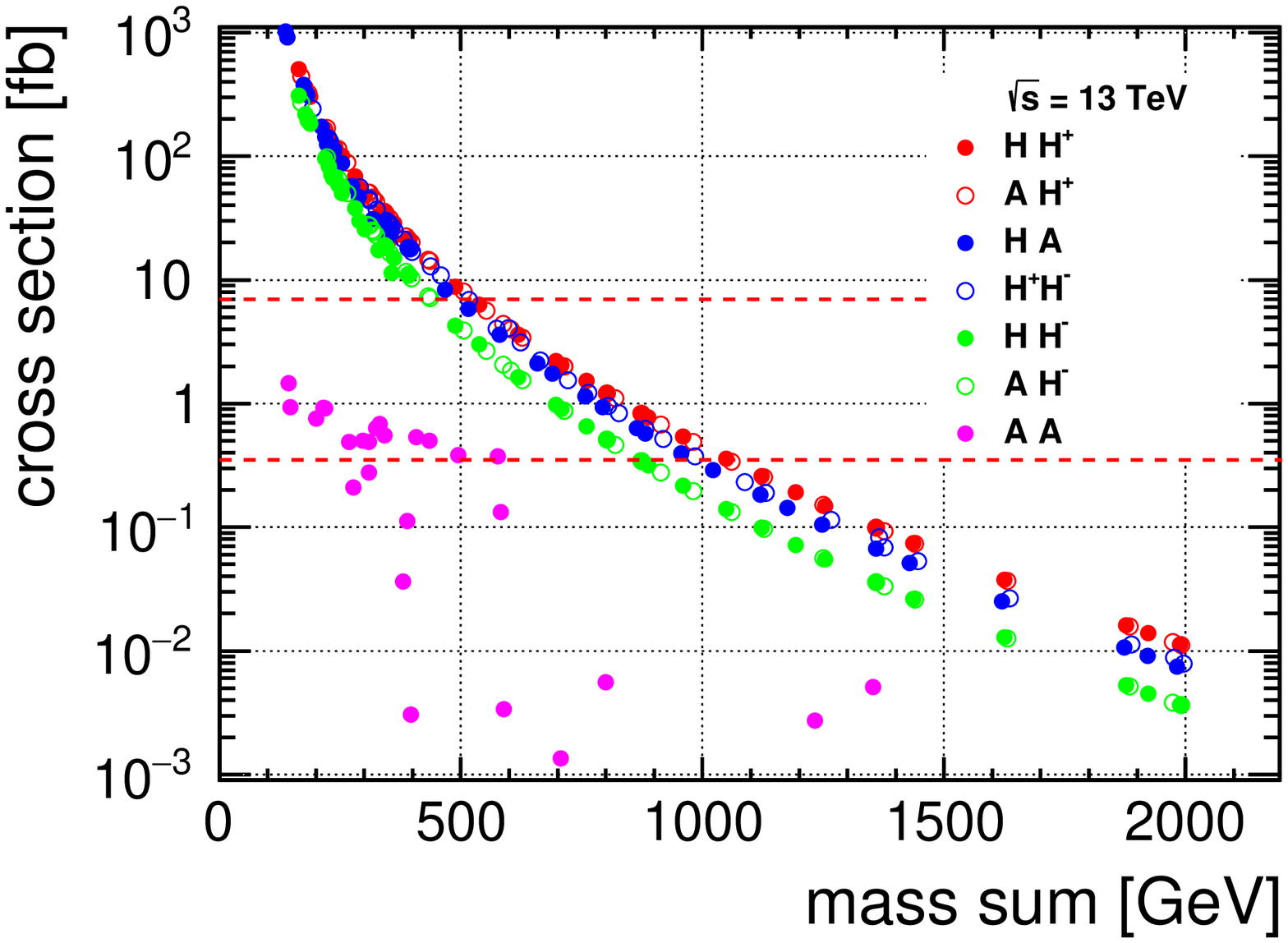}
\end{center}
\end{minipage}
\begin{minipage}{0.42\textwidth}
\begin{center}
\includegraphics[width=\textwidth]{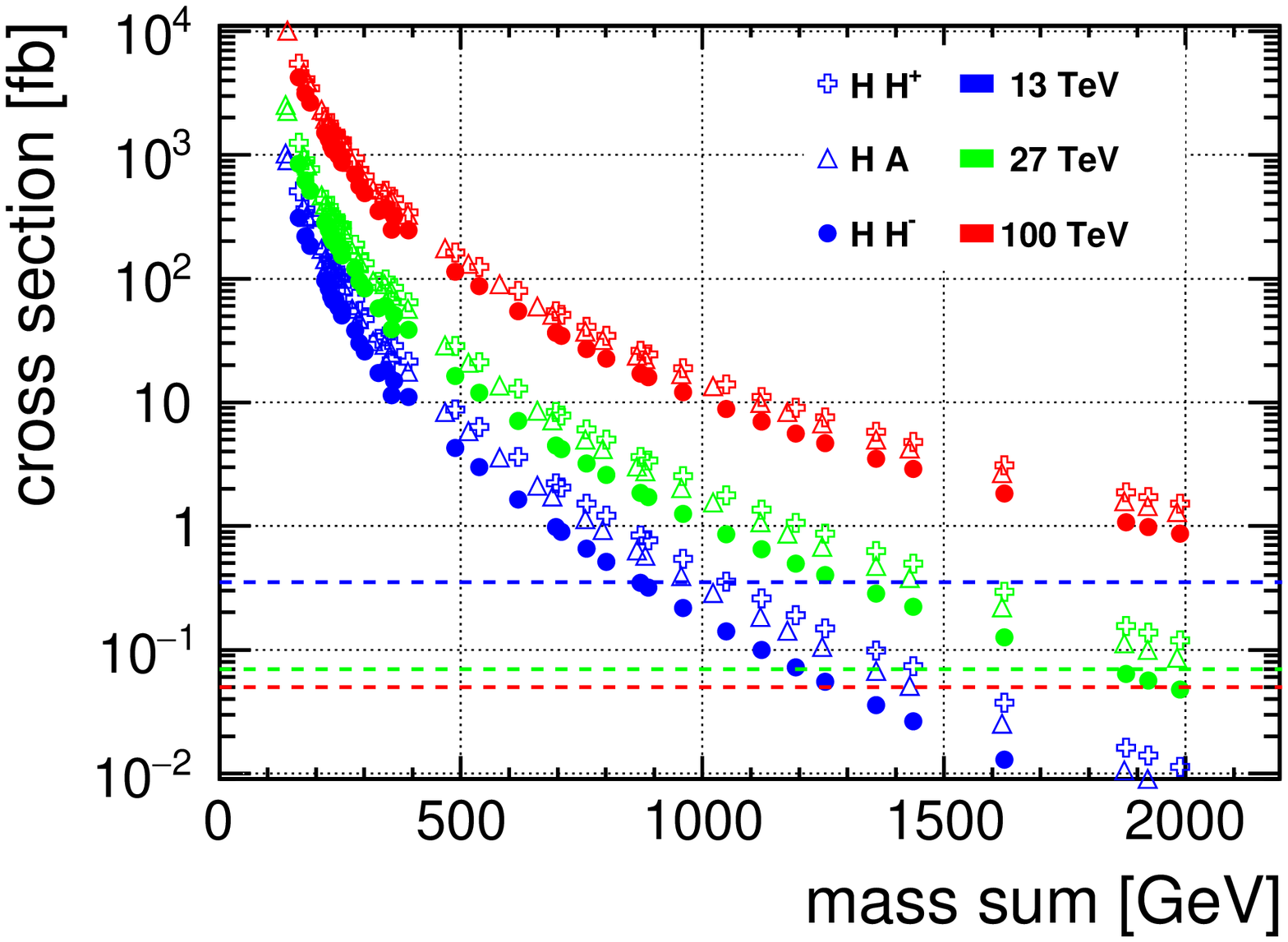}
\end{center}
\end{minipage}
\begin{minipage}{0.42\textwidth}
\begin{center}
\includegraphics[width=\textwidth]{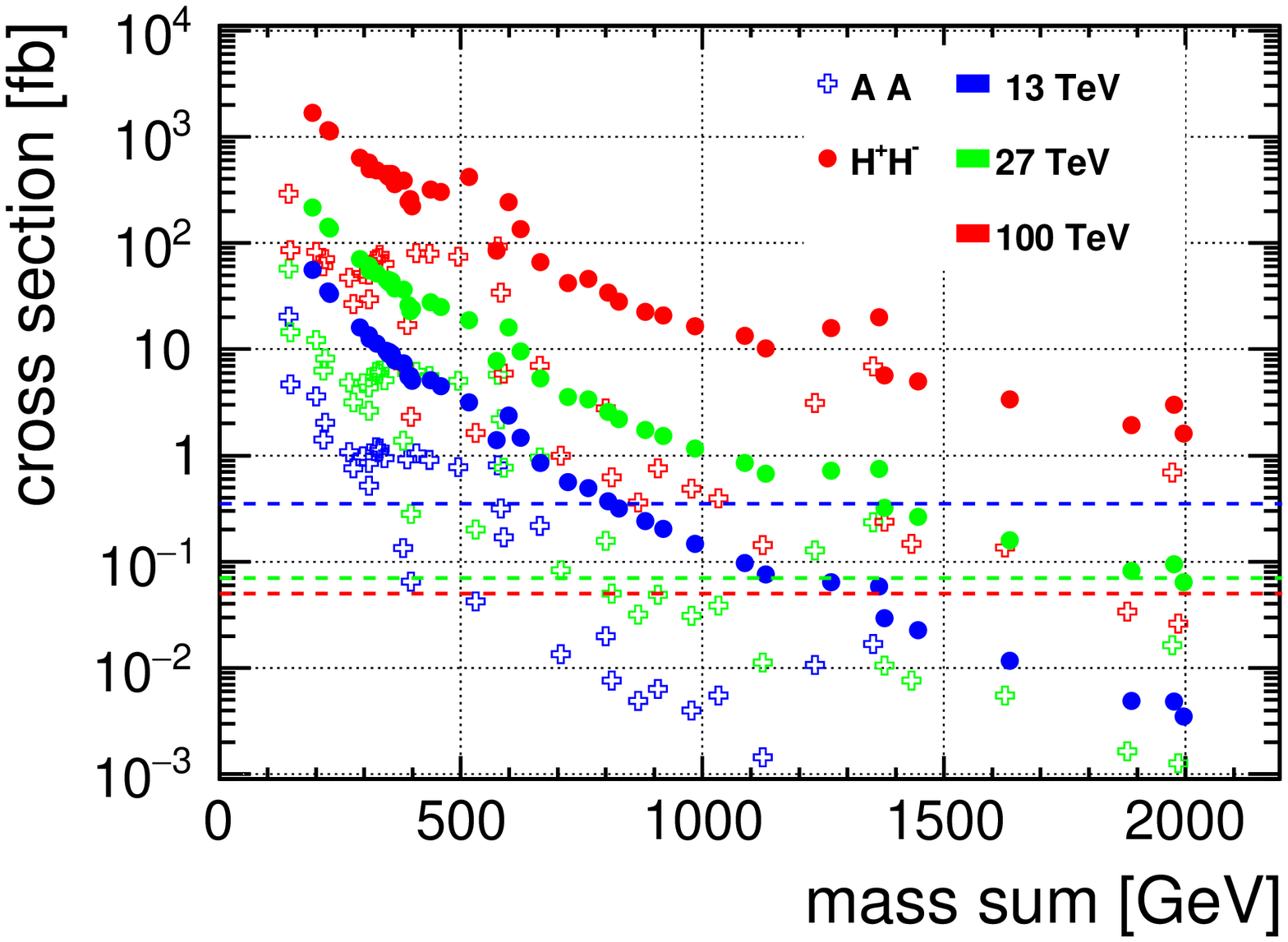}
\end{center}
\end{minipage}
\begin{minipage}{0.42\textwidth}
\begin{center}
\includegraphics[width=\textwidth]{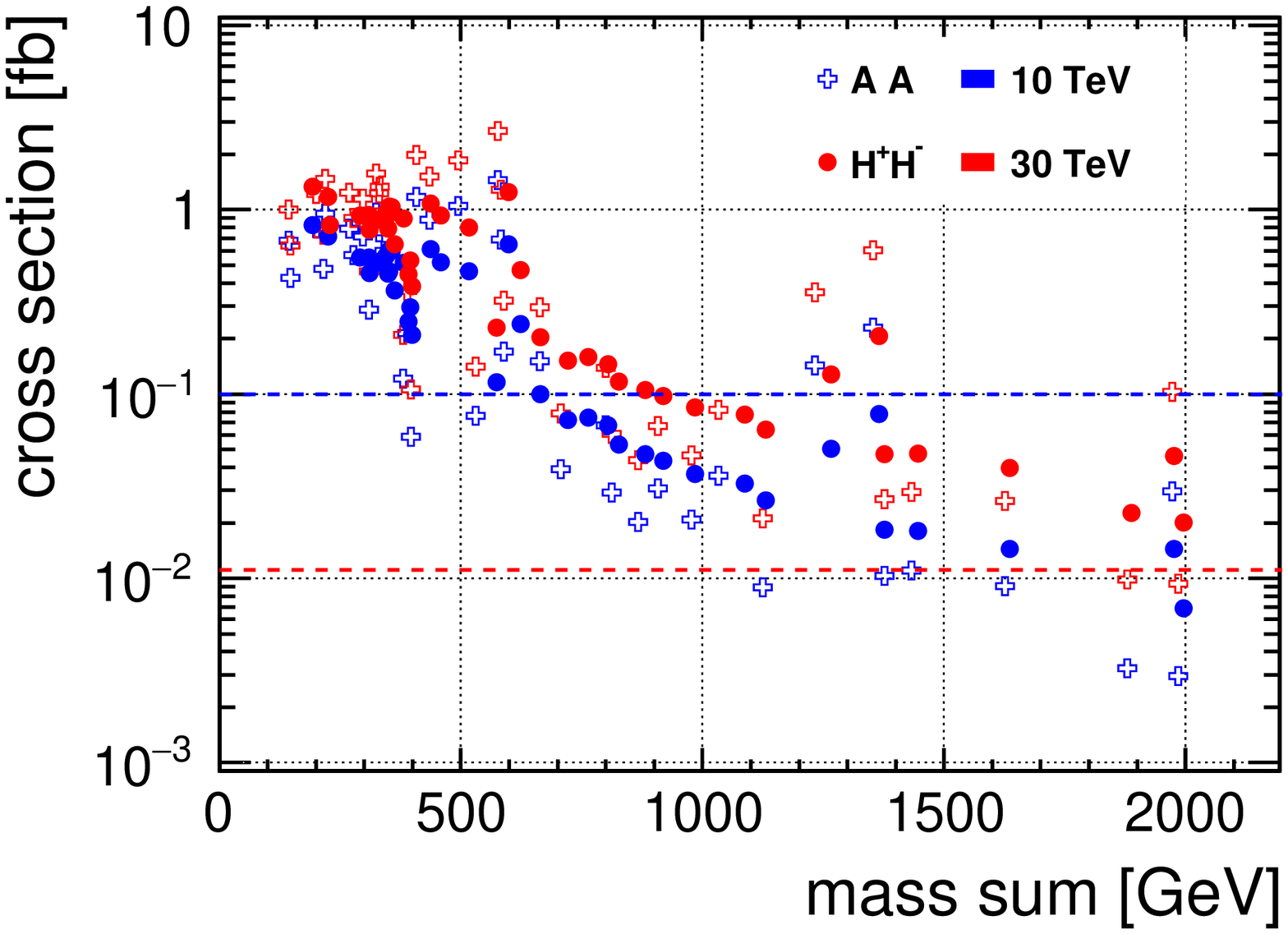}
\end{center}
\end{minipage}
\end{center}
\caption{\label{fig:idm} Predictions for production cross sections for various processes and collider options. {\sl Top left:} Predictions for various pair-production cross sections for a $pp$ collider at 13 \TeV, as a function of the mass sum of the produced particles. {\sl Top right:} Same for various center-of-mass energies. {\sl Bottom left:} VBF-type production of $AA$ and $H^+\,H^-$ at various center-of-mass energies for $pp$ colliders. {\sl Bottom right:} Same for $\mu^+\mu^-$ colliders. Taken from \cite{Kalinowski:2020rmb}. The lines correspond to the cross-sections required to prodce at least 1000 events using the respective design luminosity.}
\end{figure}
\end{center}

In \cite{Kalinowski:2018ylg} (see also \cite{Kalinowski:2018kdn}), benchmark points were suggested which cover dark scalar masses up to 1 \TeV~ and are in agreement with the current constraints discussed above. We here focus on the investigation of these points at various current and future colliders, concentrating on pair-production processes
\begin{\eqn*}
{ p\,p\,\rightarrow\,HA,\,H H^\pm,\,A H^\pm, H^+ H^-};\;\;\;
{\ell^+\,\ell^-\,\rightarrow\,HA,\,H^+ H^-}
\end{\eqn*}
for various collider options, where $\ell\,\in\,\left\{e,\,\mu\right\}$. For some of these, we also considered VBF-type production processes, where the above are augmented by 2 additional jets for $pp$ and neutrinos for $\ell\ell$ machines. The different collider specifications considered here are given in table \ref{tab:specs}, where more details can be found in \cite{Kalinowski:2020rmb}.
Figure \ref{fig:idm} shows the cross-sections for various collider options and center-of-mass energies, as a function of the sum of the two pair-produced particles. The dashed lines signify the minimal cross sections needed for a generation of 1000 events assuming collider center-of-mass energies and luminosities as specified in table \ref{tab:specs}. In summary, we can state that already at the HL-LHC, all considered processes should be accessible (via the criterium above) up to a mass sum of 2 \TeV~ apart from $AA$ pair-production. The latter process depends on the parameter 
$\bar{\lam}_{345}\,\equiv\,\lam_3+\lam_4-\lam_5\,=\,\lam_{345}-2\,\frac{M_H^2-M_A^2}{v^2}$
which determines the $hAA$ coupling, i.e. also mass-differences between $A$ and $H$ are important. Furthermore, $\lam_{345}$ is highly constrained by direct detection \cite{Ilnicka:2015jba,Ilnicka:2018def,Dercks:2018wch,Kalinowski:2018ylg,Kalinowski:2020rmb}. For $AA$ production, therefore, the mass scale, defined as the sum of produced particle masses, can be up to 600 GeV~ at the HL-LHC and 2 \TeV at the FCC-hh. At lepton colliders, on the other hand, the VBF channel is needed in order to access this process, where mass scales are around 600 \GeV/ 1.4 \TeV/ 2 \TeV~ for CLIC with a center-of-mass energy of 3 \TeV/ a muon-collider with a center-of-mass energy of 10 \TeV/ 30 \TeV, respectively.
\begin{center}
\begin{table}
\begin{center}
\begin{tabular}{c|c|c|c}
collider&cm energy [\TeV]&$\int\mathcal{L}$&1000 events [\fb]\\ \hline
HL-LHC&13/ 14&$3\,\ab^{-1}$&0.33\\
HE-LHC&27&$15\,\ab^{-1}$&0.07\\
FCC-hh&100&$20\,\ab^{-1}$&0.05\\ \hline
ee&3&$5\,\ab^{-1}$&0.2\\
$\mu\mu$&10&$10\,\ab^{-1}$&0.1\\
$\mu\mu$&30&$90\,\ab^{-1}$&0.01
\end{tabular}
\end{center}
\caption{\label{tab:specs} Parameter used for the different collider options discussed in the text. More details, including references, can be found in \cite{Kalinowski:2020rmb}.}
\end{table}
\end{center}

\section{THDMa}
The THDMa is a type II two-Higgs-doublet model that is extended by an additional pseudoscalar mixing with the "standard" pseudoscalar $A$ of the THDM. In the gauge-eigenbasis, the additional scalar serves as a portal to the dark sector, with a fermionic dark matter candidate, denoted by $\chi$. This model has e.g. been discussed in \cite{Ipek:2014gua,No:2015xqa,Goncalves:2016iyg,Bauer:2017ota,Tunney:2017yfp,LHCDarkMatterWorkingGroup:2018ufk,Robens:2021lov}. 

The model features the following particles in the scalar and dark matter sector: ${h,\,H,\,H^\pm,}\,{a,}\,{A,}\,{{\chi}}$. It depends on in total 12 additional parameters
\begin{eqnarray*}
{v,\,m_h,\,m_H,}\,{ m_a,}\,{m_A,\,m_{H^\pm},}\,{m_\chi};\;{\cos\lb \be-\al\rb,\,\tan\be,}\,{\sin\theta;\;y_\chi,}\,{\lam_3,}\,{\lam_{P_1},\,\lam_{P_2}},
\end{eqnarray*}
where $v$ and either $m_h$ or $m_H$ are fixed by current measurements in the electroweak sector. We refer the reader to the above references for more details.

\begin{center}
\begin{figure}
\begin{center}
\begin{minipage}{0.45\textwidth}
\begin{center}
\includegraphics[width=\textwidth]{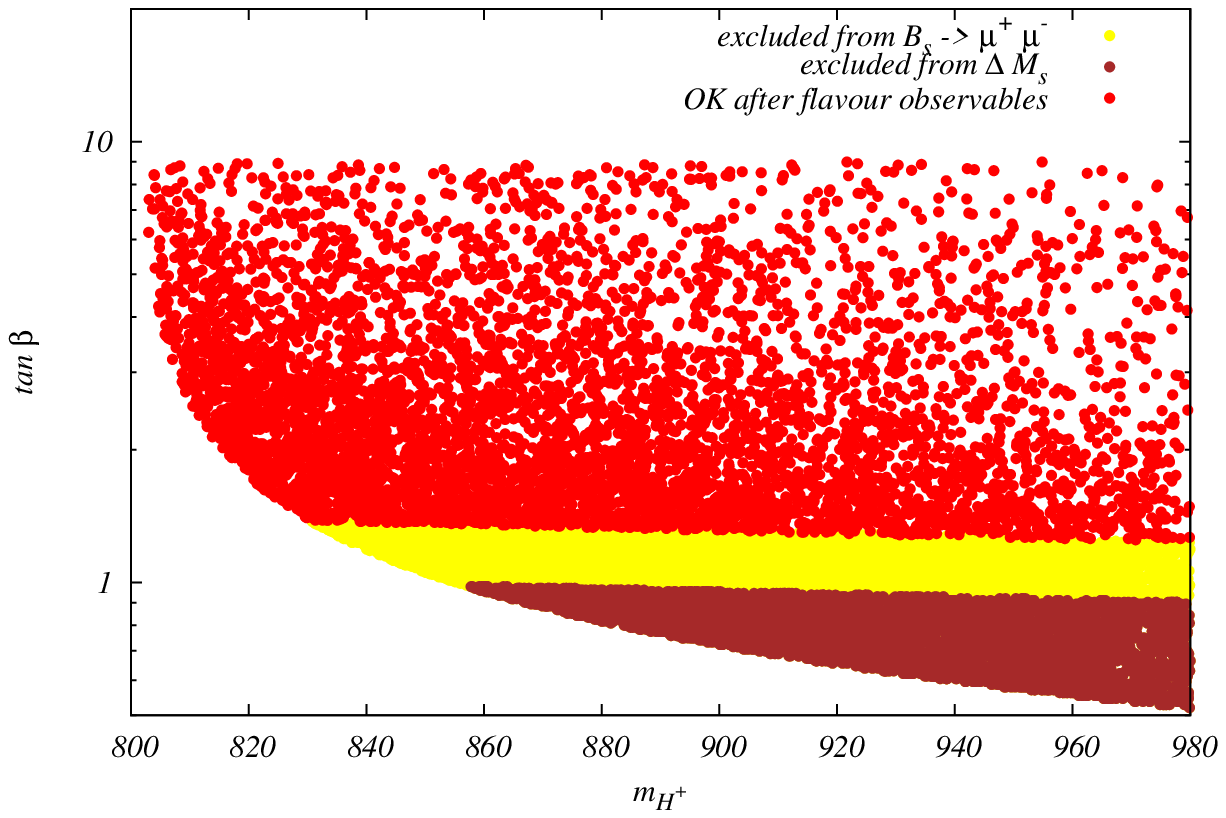}
\end{center}
\end{minipage}
\begin{minipage}{0.45\textwidth}
\begin{center}
\includegraphics[width=\textwidth]{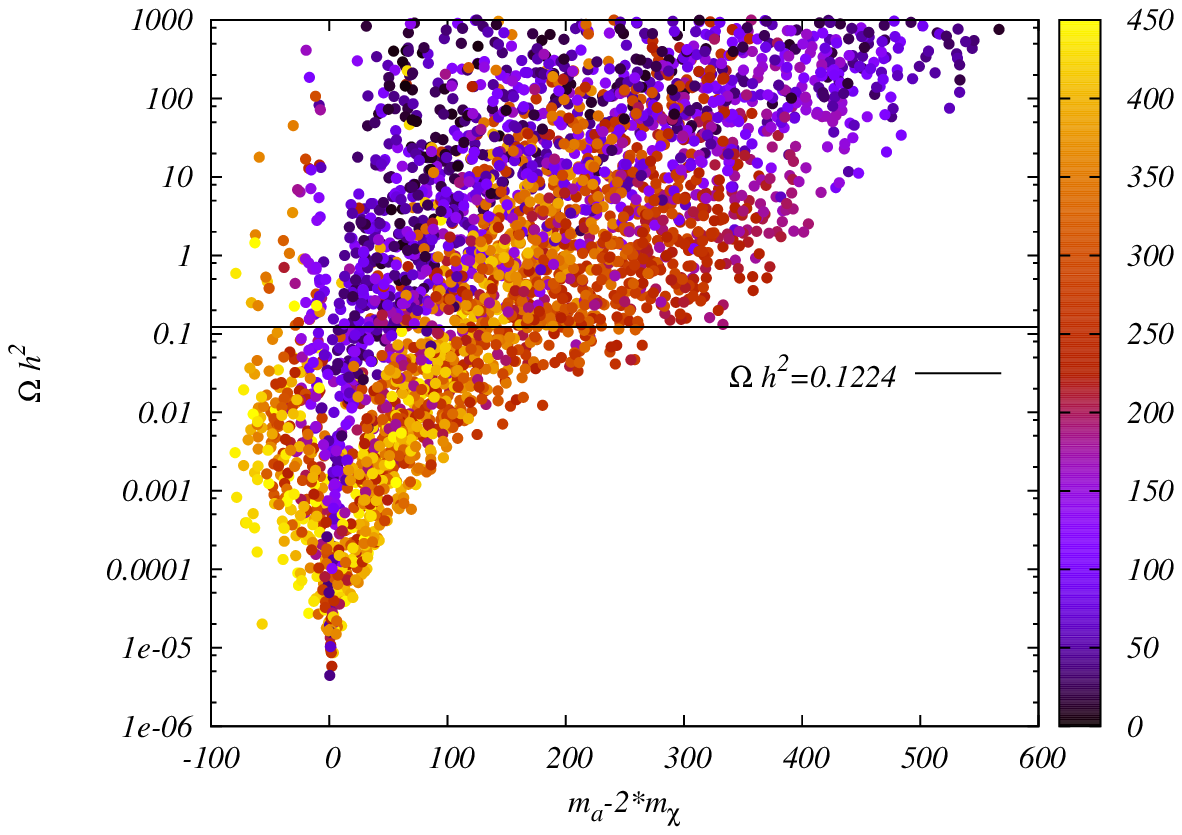}
\end{center}
\end{minipage}
\end{center}
\caption{\label{fig:thdmab} {\sl Left:} Bounds on the $\lb m_{H^\pm},\,\tan\be\rb$ plane from B-physics observables, implemented via the SPheno \cite{Porod:2011nf}/ Sarah \cite{Staub:2013tta} interface,  and compared to experimental bounds \cite{combi,Amhis:2019ckw}. The contour for low $\lb m_{H^\pm,\,\tan\be}\rb$ values stems from \cite{Misiak:2020vlo,mm}. {\sl Right:} Dark matter constraints in the THDMa model. {\sl Right:} Dark matter relic density as a function of $m_a-2\,m_\chi$, with $m_\chi$ defining the color coding. The typical resonance-enhanced relic density annihilation is clearly visible. Figures taken from \cite{Robens:2021lov}.}
\end{figure}
\end{center}

\begin{center}
\begin{figure}
\begin{center}
\begin{minipage}{0.45\textwidth}
\begin{center}
\includegraphics[width=\textwidth]{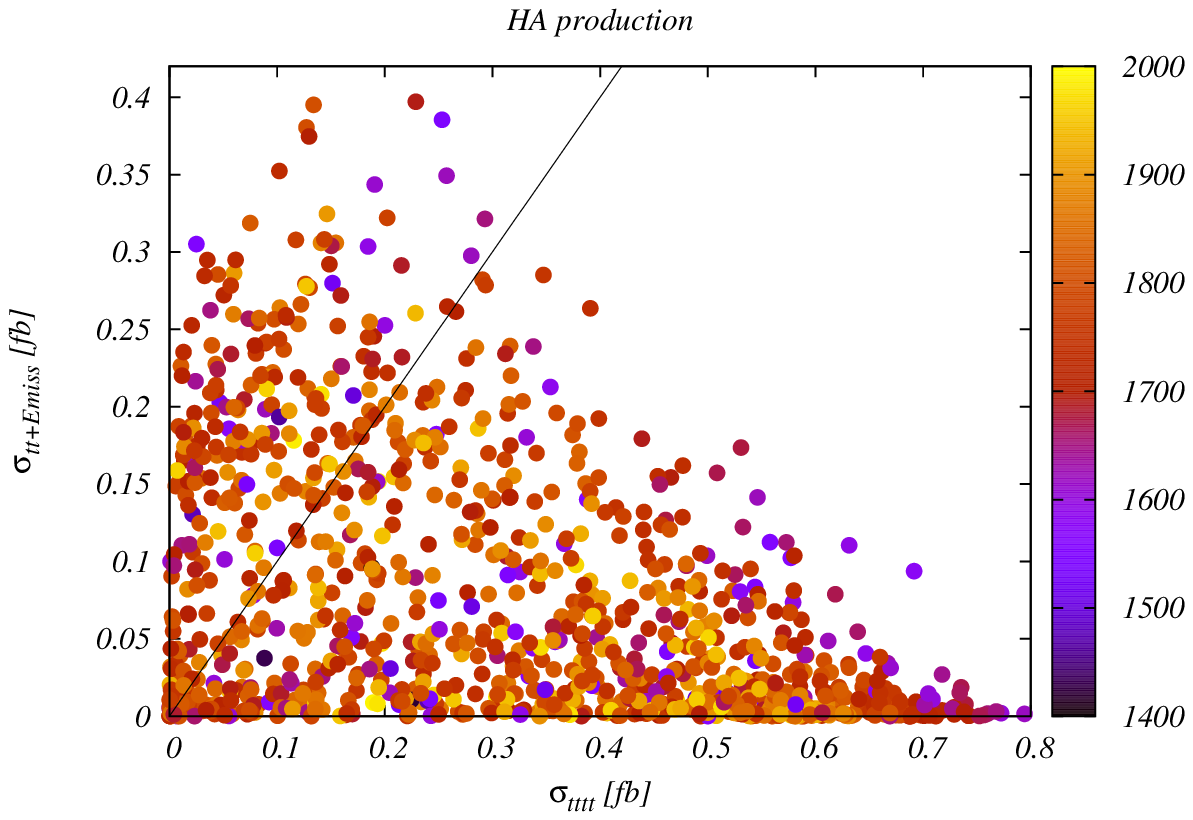}
\end{center}
\end{minipage}
\begin{minipage}{0.45\textwidth}
\begin{center}
\includegraphics[width=\textwidth]{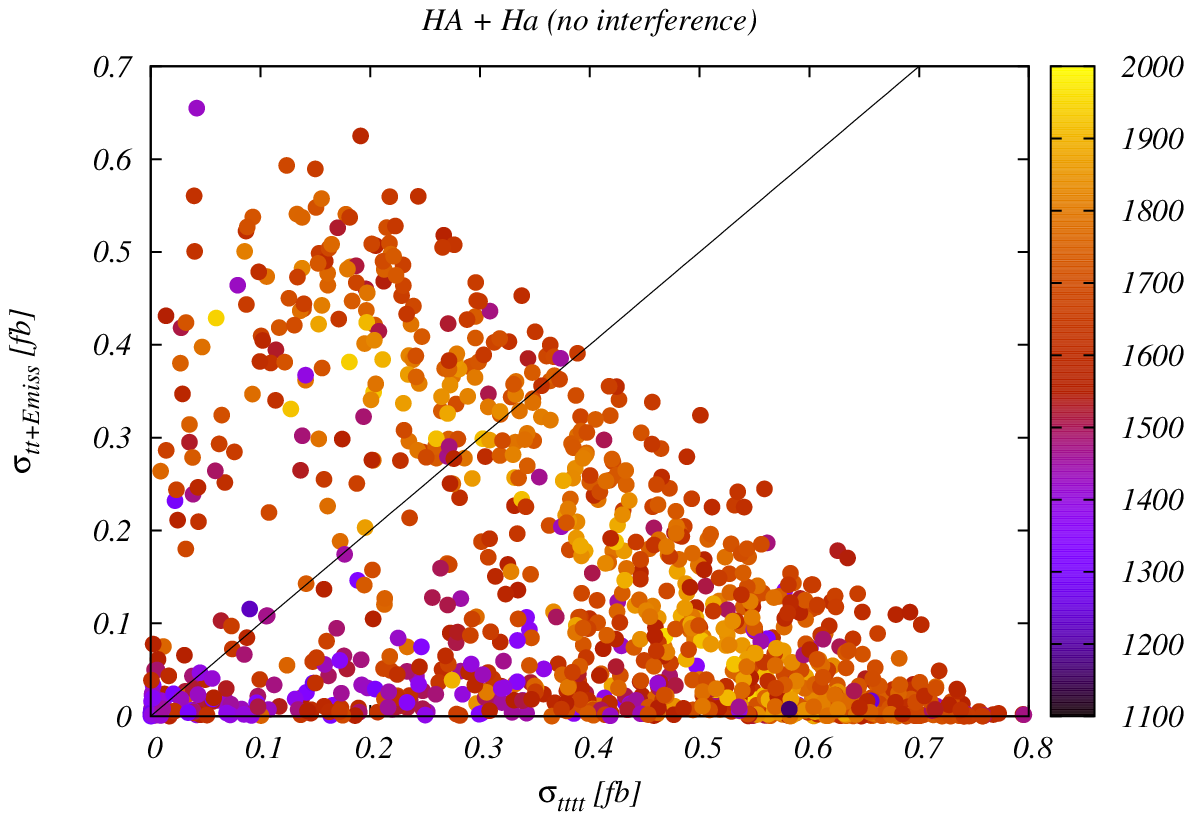}
\end{center}
\end{minipage}
\end{center}
\caption{\label{fig:thdmaatee} Production cross sections for $t\bar{t}t\bar{t}$ (x-axis) and $t\bar{t}+\slashed{E}$ (y-axis) final state in a factorized approach, for an $e^+e^-$ collider with a 3 \TeV center-of-mass energy. {\sl Left:} mediated via $HA$, {\sl right:} mediated via $HA$ and $Ha$ intermediate states. Color coding refers to $m_H+m_A$ {\sl (left)} and $M_H+0.5\times\,\lb m_A+m_a\rb$ {\sl(right)}. Figures taken from \cite{Robens:2021lov}.}
\end{figure}
\end{center}

We here report on results of a scan that allows all of the above novel parameters float in specific predefined ranges \cite{Robens:2021lov}. In such a scenario, it is not always straightforward to display bounds from specific constraints in 2-dimensional planes. Two examples where this is possible are given in figure \ref{fig:thdmab}. In the first plot, we show bounds in the $\lb m_{H^\pm},\,\tan\be \rb$ plane from B-physics observables. The result is similar in a simple THDM, and shows that in general low masses $m_{H^\pm}\lesssim\,800\,\GeV$  as well as values $\tan\be\lesssim\,1$ are excluded. The second plot features the relic density as a function of the mass difference $m_a-2\,m_\chi$. Here, a behaviour can be observed that is typical in many models with dark matter candidates: in the region where this mass difference remains small, relic density annihilates sufficiently to stay below the observed relic density bound. On the other hand, too large differences lead to values $\Omega\,h_c\,\gtrsim\,0.12$ and therefore are forbidden from dark matter considerations. Finally, we can investigate which cross-section values would still be feasible for points that fulfill all constraints \cite{Robens:2021lov} at $e^+e^-$ colliders. We are specifically interested in signatures that include missing energy and therefore do not exist in a THDM without a portal to the dark sector. Due to alignment, processes like $e^+e^-\,\rightarrow\,hA, ha$ are suppressed, which makes $e^+e^-\,\rightarrow\,HA, Ha$ the most interesting channel featuring novel signatures. Due to the interplay of B-physics and electroweak constraints, such points typically have mass scales $\gtrsim\,1\,\TeV$. We therefore consider production cross sections for an $e^+e^-$ collider with a center-of-mass energy of 3 \TeV. The corresponding production cross sections are displayed in figure \ref{fig:thdmaatee}, where we give predictions for $t\,\bar{t}\,t\,\bar{t}$ and $t\,\bar{t}+\slashed{E}$ final states using a factorized approach. We see that there is a non-negligible number of points where the second channel is dominant. A "best" point with a large rate for $t\,\bar{t}+\slashed{E}_\perp$ has been presented in \cite{Robens:2021lov}.

\section{Conclusion and Outlook}
In this work, we presented two models that extend the particle content of the SM and also provide at least one dark matter candidate. We have presented production cross sections for various standard pair-production modes within these models; for the IDM, we have given an estimate of mass range reachibility based on a simple counting criterium. A more dedicated investigation of the corresponding signatures, including background simulation and cut optimization, is in the line of future work.
\section*{Acknowledgements}
This research was supported in parts by the National Science Centre, Poland, the HARMONIA
project under contract UMO-2015/18/M/ST2/00518 (2016-2021), OPUS project under contract
UMO-2017/25/B/ST2/00496 (2018-2021), as well as the COST action CA16201 - Particleface. 


\end{document}